\documentclass[12pt,aps,floats]{revtex4}

\begin{document}
\def\bigint{{\displaystyle\int}}
\def\simlt{\stackrel{<}{{}_\sim}}
\def\simgt{\stackrel{>}{{}_\sim}}

\title{Generation of primordial cosmological density inhomogeneities with
scale invariant power spectrum during the standard radiation dominated
expansion of the universe}

\author{David H. Oaknin}
\affiliation{
Rafael, \\
Haifa 31021, Israel. \\
e-mail: davidoa@rafael.co.il}

\begin{abstract}
The most distinctive feature of the primordial density inhomogeneities that existed
in the cosmic plasma at the instant of decoupling is their scale invariant power spectrum
${\cal P}(k) \sim k$ over the range $k \ll H_{eq}$ of modes with cosmologically large
comoving wavelength. We characterize this feature in real space, in terms of their
correlation function at two points. We show that over cosmologically large comoving distances
$r \gg H^{-1}_{eq}$ the primordial inhomogeneities were (anti)correlated as $f(r) \sim - r^{-4}$. 
We revisit the so-called {\it origin of structures problem} of the standard cosmology at the 
light of this observation. Our conclusions contradict the current wisdom on this issue.
\end{abstract}

\maketitle

1. Tiny temperature anisotropies over the sky in the blackbody spectrum of the
cosmic microwave background radiation~(CMBR) were first detected by the Cosmic
Background Explorer~(COBE) at the beginning of the 90's \cite{COBE} and their
amplitude and statistical correlations were precisely measured with angular
resolution of a fraction of a degree by the Wilkinson Microwave Anisotropy
Probe~(WMAP) a decade later \cite{WMAP}. Long before, during
the 70's, these temperature anisotropies had been theoretically predicted
\cite{HarrisonZeldovich,Peebles} as the imprints of the primordial
seed of tiny density inhomogeneities that must have existed in the cosmic
plasma at the time of decoupling and from which the present large scale
structures of the universe would have developed through gravitational collapse
during the subsequent stage of matter dominated expansion.
The measurement with higher precision and better spatial resolution of the
primordial seed of cosmological density inhomogeneities, as well as the
analysis and interpretation of its statistical features, is the most crucial
enterprise in modern cosmology \cite{www.WMAP.site}, because
of the wealth of information that it can disclose on the geometry and content of
the universe, as well as on the physical mechanisms that operated in it during
its early history.

The available observational data on this random seed of primordial cosmological
density inhomogeneities is compatible with the assumption of gaussianity
(see, for example, \cite{Dennis}) and, therefore, it is generally accepted that
their statistical correlations at two points $f({\vec x}, {\vec y})$ completely
describe them. It is also generally assumed, along the fundamental cosmological
principle of standard Friedmann-Robertson-Walker~(FRW) cosmology, that these
inhomogeneities are statistically homogeneous and isotropic. Under these
assumptions the statistical correlations at two points are a function only of the
comoving distance between them, $f({\vec x}, {\vec y}) = f(|{\vec x}-{\vec y}|)$.
It is quite common in the literature to represent these correlations in terms of
their power spectrum in Fourier space (see, for example, \cite{BrandenbergerMukhanovFeldman}),

\begin{equation}
\label{powerspectrum}
{\cal P}(k) = \frac{1}{\rho^2_0}\ \int d^3{\vec r}\ f(r)\ e^{-i{\vec k}\cdot {\vec r}} =
\frac{4\pi}{\rho^2_0}\ \int_0^{\infty} dr\ r^2\ f(r)\ \frac{sin(k r)}{k r}.
\end{equation}
The power spectrum has units of volume and is necessarily a non-negative function,
${\cal P}(k) \ge 0$. The normalization factor $\rho_0$ in this definition is
the average cosmic density per unit comoving volume. For the sake of simplicity, we use
throughout this paper comoving coordinates defined with the scale factor of the universe
at decoupling as reference, $a(t_{eq})=1$.


2. The most characteristic statistical feature of the primordial seed of
cosmological density inhomogeneities that existed in the cosmic plasma at the time of
decoupling is the scale invariance of their power spectrum in the range of
wavenumbers with comoving wavelength much larger than the comoving size of the
FRW causal horizon at that instant $H^{-1}_{eq}$,

\begin{equation}
\label{scaleinvariance}
{\cal P}(k) = {\cal O}(k), \hspace{0.3in} at \hspace{0.3in} k \ll H_{eq}.
\end{equation}
The notation ${\cal O}(k^p)$ used here is borrowed from calculus
to denote an infinitesimal function of order $p$. Scale invariance means $p = 1$,
that is, the power spectrum of the primordial density inhomogeneities is an
infinitesimal of linear order in the range of modes with cosmologically large
comoving wavelength. Let us remind that the comoving size of the FRW causal
horizon at decoupling $H^{-1}_{eq}$ is roughly two orders of magnitude shorter
than the comoving size $H^{-1}_0$ of the causal horizon at present,
$H^{-1}_0 \simeq 100\ H^{-1}_{eq}$.

Scale invariance, which was theoretically predicted as early as
the the very existence of the primordial seed of cosmological density inhomogeneities
in the radiation dominated  cosmic plasma \cite{HarrisonZeldovich}, has been
observationally confirmed by the data collected by the WMAP \cite{WMAP}.
Nonetheless, this feature (\ref{scaleinvariance}) still poses a major theoretical
challenge for the standard FRW cosmological framework:
widely reported analysis \cite{wisdom} seem to indicate that the power spectrum of the
primordial density inhomogeneities that could have been generated in the radiation dominated
cosmic plasma by standard physical mechanisms should be suppressed in the cosmological
range $k \ll H_{eq}$ much stronger than linearly. Indeed, as ${\cal P}(k) = {\cal O}(k^4)$
an infinitesimal of, at least, fourth order.

The theoretical reasoning behind this claim goes as follows. First, it is argued
that in the framework of standard FRW cosmology causal constrains forbid statistical
correlations over comoving distances longer than the horizon and, hence, the power
spectrum in the range $k \ll H^{-1}_{eq}$ can be estimated by Taylor expanding the basis of
Fourier functions in (\ref{powerspectrum}) around $k = 0$,

\begin{eqnarray}
\label{Taylor}
{\cal P}(k) = \frac{4\pi}{\rho^2_0}\ \int_0^{\infty} dr\ r^2\ f(r) \left(1 - \frac{(k r)^2}{3!} + {\cal O}((k r)^4)\right).
\end{eqnarray}
The spatial integral is then moved rightward before the Taylor summation to
get,

\begin{equation}
\label{expansion}
{\cal P}(k) = \frac{4\pi}{\rho^2_0} \left[\int_0^{\infty} dr\ r^2\ f(r)\right] - \frac{4\pi}{3! \rho^2_0} \left[\int_0^{\infty} dr\ r^4\ f(r)\right] k^2 + {\cal O}(k^4).
\end{equation}
As the random density inhomogeneities are constrained to preserve the total
mass (energy),

\begin{equation}
\label{mass_constrain}
4\pi \int_0^{\infty} dr\ r^2\ f(r)\ = 0,
\end{equation}
as well as the moment of inertia of the distribution,

\begin{equation}
\label{inertial_constrain}
4\pi \int_0^{\infty} dr\ r^4\ f(r) = 0,
\end{equation}
it is straightforward to conclude that their power spectrum must be an
infinitesimal of, at least, fourth order in the cosmological range,

\begin{equation}
\label{quartic}
{\cal P}(k) = {\cal O}(k^4) \hspace{0.3in} at \hspace{0.3in} k \ll H_{eq}.
\end{equation}
The discrepancy between the scale invariant power spectrum of the actual primordial
cosmological density inhomogeneities (\ref{scaleinvariance}) and the widely reported
theoretical prediction (\ref{quartic}) is commonly known in the literature as
{\it the origin of structures problem} of standard FRW cosmology. The widely accepted
conclusion inferred from this disagreement is that the scale invariant power spectrum
(\ref{scaleinvariance}) of the primordial cosmological density inhomogeneities is a
pristine signature of an early period of inflationary cosmology, which cannot have been
altered later on during the subsequent stage of radiation dominated expansion of the universe
(see, for example, \cite{Turner}). In consequence, most efforts in theoretical cosmology
during the thirty years that have passed since this argument was first laid down have been
directed towards the implementation of the inflationary paradigm \cite{Linde} in a theoretically
consistent cosmological framework compatible with the available observational data \cite{Tegmark}.


3. The motivation of this paper is rather different. We intend to test an implicit
and unadverted assumption in the reasoning that leads to the statement of the
{\it origin of structures problem} of standard FRW cosmology. In particular,
we notice that in order to justify expansion (\ref{expansion}) it is crucial to assume
$f(r)\ \simeq\ 0$\  for\  $r\ >\ H^{-1}_{eq}$, that causal constrains in the standard FRW
cosmological framework force the two points correlation function to cancel or, more
precisely, to be strongly suppressed at comoving distances larger than the causal horizon.
Indeed, if the two points correlation function has compact support it is
obviously legitimate to perform the spatial integral prior to the Taylor
summation.

The permutation in the order of performing these two operations is not
legitimate, notwithstanding, if the suppression of the two points correlation
function at distances longer than the horizon is not strong enough. This can be readily seen from
the simple fact that the global conservation constrain (\ref{mass_constrain}),
which forces the two points correlation function to be suppressed as
$f(r) \sim {\it o}(1/r^3)$ as $r \rightarrow \infty$, guarantees that expression (\ref{powerspectrum}) is properly defined, but it does not guarantee that expression (\ref{expansion}) is also proper.

This observation is pertinent because causal constrains in the standard FRW
cosmology do not necessarily force that the statistical correlations over
comoving distances larger than the causal horizon do vanish.
Actually, the global constrain (\ref{mass_constrain}) can force
$f(r)\ \neq\ 0$\ for\ $r\ >\ H^{-1}_{eq}$,
non-zero statistical correlations in the radiation dominated cosmic plasma
over comoving distances larger than the causal horizon. This can be readily
seen from the following simple consideration. We notice that the radiation
dominated cosmic plasma before decoupling supports the propagation of weak
shock waves that travel with sound velocity $c_s = 1/\sqrt{3}$ (in natural
units). These waves erase all density fluctuations with comoving wavelength
shorter than the sound horizon and, hence, force positive statistical
correlations between comoving sites separated by shorter distances. That is,
$f(r)\ >\ 0$\ for\ $r\ <\ H^{-1}_{eq}$, which is obviously not compatible with the demand
that statistical correlations do vanish over comoving distances longer than the horizon,
since the global conservation constrain (\ref{mass_constrain}) must be fulfilled.

Let us remark at this point that
$\int dr\ r^2\ f(r) = 0 \hspace{0.03in} \Longleftrightarrow \hspace{0.03in} lim_{k \rightarrow 0}\ {\cal P}(k) = 0$
and, therefore, the global constrain (\ref{mass_constrain}) is necessarily fulfilled in
any causally constrained cosmological framework.


4. In order to perform a careful estimation of the limiting behavior of the
power spectrum (\ref{powerspectrum}) in modes $k \ll H_{eq}$ for two points
correlation functions that, in general, have non-compact support we separate,

\begin{equation}
\label{twoterms}
{\cal P}(k) = {\cal P}_{short}(k) + {\cal P}_{long}(k),
\end{equation}
the contributions of statistical correlations over comoving distances shorter
and larger than the causal horizon,

\begin{equation}
\label{short}
{\cal P}_{short}(k) = \frac{4\pi}{\rho^2_0} \int_0^{H^{-1}_{eq}} dr\ r^2\ f_{short}(r)\
\frac{sin(k r)}{k r},
\end{equation}

\begin{equation}
\label{long}
{\cal P}_{long}(k) = \frac{4\pi}{\rho^2_0} \int_{H^{-1}_{eq}}^{\infty} dr\ r^2\ f_{long}(r)\
\frac{sin(k r)}{k r}.
\end{equation}
We have chosen to denote the projection of the two points correlation function
$f(r)$ upon the sphere of causally connected comoving sites with the label {\it short},
$f_{short}(r) \equiv f(r)|_{r \in [0, H^{-1}_{eq}]}$, and the projection upon
its complementary volume with the label {\it long}, $f_{long}(r) \equiv
f(r)|_{r \in [H^{-1}_{eq}, \infty)}$. As we have noticed above, $f_{long}(r) \sim
{\it o}(1/r^3)$ as $r \rightarrow \infty$ for the constrain
(\ref{mass_constrain}) to be properly defined.

The contribution (\ref{short}) of the two points correlations between causally connected
comoving sites can be safely estimated in the range of interest $k \ll H_{eq}$ by Taylor
expanding the Fourier functions around $k = 0$, as we did in (\ref{expansion}), ${\cal P}_{short}(k) = \frac{4\pi}{\rho^2_0} \left[\int_0^{H^{-1}_{eq}} dr\ r^2\
f_{short}(r)\right] - \frac{4\pi}{3! \rho^2_0}  \left[\int_0^{H^{-1}_{eq}} dr\
r^4\ f_{short}(r)\right] k^2+ {\it O}(k^4)$.  We then use the global constrain
(\ref{mass_constrain}) to write the first term in the right hand side of the last expression
as a contribution from statistical correlations over comoving distances longer than the horizon,
$\int_0^{H^{-1}_{eq}} dr\ r^2\ f_{short}(r) = -\int_{H^{-1}_{eq}}^{\infty} dr\ r^2\ f_{long}(r)$,
and get that the total power spectrum (\ref{twoterms}) in the range of very
small wavenumbers $k \ll H_{eq}$ is,

\begin{eqnarray}
\nonumber
{\cal P}(k) & = & \frac{4\pi}{\rho^2_0} \int_{H^{-1}_{eq}}^{\infty} dr\ r^2\ f_{long}(r)\
\left(\frac{sin(k r)}{k r} - 1\right) \\
\label{emet} & & - \frac{4\pi}{3! \rho^2_0} \left[\int_0^{H^{-1}_{eq}} dr\ r^4\
f_{short}(r)\right] k^2 + {\cal O}(k^4).
\end{eqnarray}
The object of interest of our revised analysis is the contribution
of the two points correlations over long distances,

\begin{equation}
\label{truth}
{\cal P}_*(k) = \frac{4\pi}{\rho^2_0} \int_{H^{-1}_{eq}}^{\infty} dr\ r^2\ f_{long}(r)\
\left(\frac{sin(k r)}{k r} - 1\right).
\end{equation}
If these correlations are strongly suppressed this contribution is, not surprisingly,
an infinitesimal of, at least, second order. More precisely,
$f_{long}(r) = {\cal O}(1/r^{n}),\ n \ge 5, \hspace{0.05in} as \hspace{0.05in}
r \rightarrow \infty \hspace{0.1in} \Longrightarrow \hspace{0.1in}
{\cal P}_*(k) = {\cal O}(k^2), \hspace{0.05in} for \hspace{0.05in} k
\ll H_{eq}$.
On the other hand, when the two points correlations at long distances are only
mildly suppressed,

\begin{equation}
\label{the_law}
f_{long}(r) \simeq -{\cal C}\ \rho^2_0\ H^{-(4 + \epsilon)}_{eq}\ r^{-(4+\epsilon)},
\hspace{0.4in}
\epsilon \in (-1,+1),
\end{equation}
where ${\cal C}$ is some dimensionless constant, their contribution to the
power spectrum in modes with cosmologically large comoving wavelength
is an infinitesimal of order lower than two,

\begin{equation}
\label{final}
{\cal P}_*(k) = -4\pi\ {\cal C}\ H^{-(4+\epsilon)}_{eq}\
\left[\int_{k H^{-1}_{eq}}^{\infty} dx\ \frac{1}{x^{2+\epsilon}}
\left(\frac{sin(x)}{x} - 1\right)\right] k^{1+\epsilon} =
{\cal O}(k^{1+\epsilon}),
\end{equation}
because as $k \rightarrow 0$ the integral within the squared brackets tends to a finite
negative constant, $-\infty < {\cal A} \equiv lim_{X \rightarrow 0} \int_{X}^{\infty} dx\
\frac{1}{x^{2+\epsilon}} \left(\frac{sin(x)}{x} - 1\right) < 0$.
Actually, the correlations over long distances (\ref{the_law}) give a scale invariant
contribution to the total power spectrum (\ref{emet}) if $\epsilon \simeq 0$,

\begin{equation}
\label{gettotal}
{\cal P}(k) \simeq - 4\ \pi\ {\cal C}\ H^{-4}_{eq}\ {\cal A}\
k - \frac{4\pi}{3! \rho^2_0} \left[\int_0^{H^{-1}_{eq}} dr\ r^4\
f_{short}(r)\right] k^2 + {\it O}(k^4).
\end{equation}
That is,

\begin{eqnarray}
\nonumber
\int_0^{\infty} dr\ r^2\ f(r) = 0 \hspace{0.05in} and \hspace{0.05in} f(r) =
{\cal O}(1/r^{4}), \hspace{0.0in} as \hspace{0.05in} r \rightarrow \infty \hspace{0.1in}
\Leftrightarrow  \hspace{0.1in} {\cal P}(k) = {\cal O}(k), \hspace{0.1in} for \hspace{0.1in} k \ll H_{eq}
\end{eqnarray}

This derivation proves, as a side result, that scale invariant density
fluctuations (\ref{the_law}) are necessarily anticorrelated  at long
distances, ${\cal C} > 0$, because the power spectrum (\ref{gettotal}) must
be non-negative everywhere, ${\cal P}(k) \ge 0$. Actually, the global constrain
(\ref{mass_constrain}) imposes that
$\int_0^{\infty} dr\ r^2\ f(r) = \int_0^{H^{-1}_{eq}} dr\ r^2\ f_{short}(r)
- {\cal C}\ \rho^2_0\ H^{-4}_{eq}\ \int_{H^{-1}_{eq}}^{\infty} dr\ r^{-2} = \\
= \int_0^{H^{-1}_{eq}} dr\ r^2\ f_{short}(r) - {\cal C}\ \rho^2_0\ H^{-3}_{eq} = 0$,
and, therefore,

\begin{equation}
{\cal C} = \frac{H^{3}_{eq}}{\rho^2_0} \int_0^{H^{-1}_{eq}} dr\ r^2\ f_{short}(r) = \frac{1}{3 \rho^2_0}\ {\widetilde f}(0).
\end{equation}
That is, the slope of the scale invariant power spectrum (\ref{gettotal}) is fixed by
the average correlations within the causal sphere ${\widetilde f}(0) \equiv 3\ H^3_{eq}\
\int_0^{H^{-1}_{eq}} dr\ r^2\ f_{short}(r)$ and the size of the causal horizon at decoupling,
${\cal P}(k) \simeq - \frac{4\pi}{3 \rho^2_0}\ {\cal A}\ {\widetilde f}(0)\ H^{-4}_{eq}\ k$\
for\ $k \ll H_{eq}$.


5. This revised analysis naturally lead us to wonder if causal mechanisms operating in the
cosmic plasma during the radiation dominated expansion of the universe could generate 
mildly suppressed $f(r) \sim -r^{-4}$ energy density (anti)correlations over comoving distances 
longer than the causal horizon. A simple toy model shall convince the reader that
this is indeed feasible.

We consider a homogeneous and isotropic fluid that fills the infinite Minkowski space and study its
response to a local density inhomogeneity generated within it. We describe the fluid dynamics within the 
framework of relativistic hydrodynamics in order to carefully account for causality constrains. 
In this framework the flow equations are

\begin{equation}
\label{hydroEq_start}
\partial_{\mu} T^{\mu \nu}(x) = 0,
\end{equation}
where

\begin{equation}
\label{energy-momentum}
T^{\mu \nu}(x) = \left[\varepsilon(x) + P(x)\right] u^{\mu}(x) u^{\nu}(x) - P(x) g^{\mu \nu},
\end{equation}
is the isentropic energy-momentum tensor and $g^{\mu \nu} = diag(+1,-1,-1,-1)$ is the Minkowski metric tensor.
The fluid four-velocity $u^{\mu}(x) = \left(1,{\vec v}({\vec x},t)\right)/\sqrt{1-|{\vec v}({\vec x},t)|^2}$ transforms as a Lorentz vector and fulfills the normalization equation $u_{\mu}(x) u^{\mu}(x) = 1$. Finally, $\varepsilon(x)$ and $P(x)$ are, respectively, the fluid internal energy and pressure in its proper local frame. They are, obviously, invariant under Lorentz transformations.

We assume that these two thermodynamic quantities are related by an equation of state of the kind,

\begin{equation}
P(x) = c^2_s\ \varepsilon(x),
\end{equation}
where $0 < c_s < 1$ is the sound velocity in the fluid. We plug this relationship into (\ref{energy-momentum})

\begin{equation}
\label{energy-momentum2}
T^{\mu \nu}(x) = \left(1 + c^2_s\right) \varepsilon(x) u^{\mu}(x) u^{\nu}(x) - c^2_s \varepsilon(x) g^{\mu \nu},
\end{equation}
and write down the flow equations

\begin{eqnarray}
\label{hydroEqB}
u^{\mu}(x) u^{\nu}(x) \partial_{\mu} \varepsilon(x) + \varepsilon(x) \left[u^{\mu}(x) \partial_{\mu} u^{\nu}(x)
+ u^{\nu}(x) \partial_{\mu} u^{\mu}(x)\right]
- \frac{c^2_s}{\left(1 + c^2_s\right)} g^{\mu \nu} \partial_{\mu} \varepsilon(x) = 0.
\end{eqnarray}
We find convenient to formulate them as follows. First, we multiply (\ref{hydroEqB}) by $u_{\nu}(x)$ to obtain the
scalar equation

\begin{eqnarray}
\label{hydroEq0}
u^{\mu}(x) \partial_{\mu} \varepsilon(x) + \left(1 + c^2_s\right)\varepsilon(x) \partial_{\mu} u^{\mu}(x) = 0,
\end{eqnarray}
which we plug into (\ref{hydroEqB}) to find

\begin{eqnarray}
\label{hydroEqC}
\varepsilon(x) u^{\mu}(x) \partial_{\mu} u^{\nu}(x)
+ \frac{c^2_s}{1 + c^2_s} \left[u^{\mu}(x) u^{\nu}(x)
- g^{\mu \nu}\right] \partial_{\mu} \varepsilon(x) = 0.
\end{eqnarray} 
We now specialize these fully covariant equations to the preferred frame located at the origin of a spherically symmetric flow, $\varepsilon({\vec x},t) = \varepsilon(r,t)$, ${\vec v}({\vec x},t) = v(r,t) {\hat r}$. As we said, 
we assume that this flow is the response to a local density inhomogeneity generated at the origin. We find 

\begin{eqnarray}
\nonumber
\partial_t \varepsilon(r,t) + v(r,t) \partial_r \varepsilon(r,t) + \left(1 + c^2_s\right) \varepsilon(r,t)
\left[\frac{v(r,t) \partial_t v(r,t) + \partial_r v(r,t)}{1 - v^2(r,t)} + \frac{2}{r} v(r,t)\right] = 0,
\end{eqnarray}

\begin{eqnarray}
\nonumber
\varepsilon(r,t) \frac{\partial_t v(r,t) + v(r,t) \partial_r v(r,t)}{1- v^2(r,t)} + \frac{c^2_s}{1 + c^2_s} \left[v(r,t) \partial_t \varepsilon(r,t) + \partial_r \varepsilon(r,t)\right] = 0.
\end{eqnarray}
This system of hyperbolic equations can be numerically solved, in general, by the method of the characteristics \cite{Courant}. Its solutions show \cite{oaknin} that the radial flow can extend well beyond the causal light-cone of the origin. This can be easily seen, in particular, when the flow is stationary:

\begin{eqnarray}
\label{stat1}
v(r) \partial_r \varepsilon(r) + \left(1 + c^2_s\right) \varepsilon(r)
\left[\frac{\partial_r v(r)}{1 - v^2(r)} + \frac{2}{r} v(r)\right] = 0,
\end{eqnarray}

\begin{eqnarray}
\label{stat2}
\varepsilon(r) \frac{v(r) \partial_r v(r)}{1- v^2(r)} + \frac{c^2_s}{1 + c^2_s} \partial_r \varepsilon(r)
= 0.
\end{eqnarray}
It is then straightforward to find

\begin{eqnarray}
\label{incompressible_flow}
\left(\frac{v^2(r)}{c^2_s} - 1\right) \frac{\partial_r v(r)}{1- v^2(r)} -
\frac{2}{r} v(r) = 0.
\end{eqnarray}
According to this equation the flow velocity $v(r)$ everywhere beyond the light-cone is fixed by the flow velocity $v_R$ at the light-cone surface. In particular, when $v^2 \ll c^2_s < 1$ the flow is roughly
incompressible\footnote{Indeed, if $c_s=1$ the flow is perfectly incompressible.}

\begin{equation}
v(r) \sim v_R \left(\frac{R}{r}\right)^2
\end{equation}
The energy density associated to this flow is given, in this asymptotic limit, by

\begin{equation}
T^{0 0}(r) \sim \varepsilon(r) \sim \varepsilon_R\ e^{\frac{1 + c^2_s}{2 c^2_s} v^2_R \left(1 - \frac{R^4}{r^4}\right)} \sim \varepsilon_{\infty} \cdot\left(1 - \frac{1 + c^2_s}{2 c^2_s} v^2_R \frac{R^4}{r^4}\right).
\end{equation}
That is, the response of a fluid to a local density inhomogeneity can generate scale invariant
density (anti)correlations over scales well beyond its light-cone. 

These conclusions challenge the prevalent wisdom in cosmology, according to which the scale invariant power spectrum of the primordial cosmological density inhomogeneities is a pristine signature of the very early history of the universe that could have not been altered during the subsequent FRW standard expansion.

\end{document}